\documentstyle[11pt,aaspp4,tighten,flushrt]{article}

\newcommand{\kms}{\mbox{${\rm\,km\,s}^{-1}$}}
\newcommand{\pc}{\mbox{$\rm\,pc$}}

\newcommand{\lsun}{\mbox{$\,L_\odot$}}

\newcommand{\beq}{\begin{equation}}
\newcommand{\eeq}{\end{equation}}
\newcommand{\hst}{{\it HST\/}}

\lefthead{Statler}
\righthead{Nuclear Rotation Curve of M31}

\begin{document}

\title{On the Nuclear Rotation Curve of M31}

\author{Thomas S. Statler}
\affil{Department of Physics and Astronomy, 251B Clippinger Research
Laboratories, Ohio University, Athens, OH 45701, USA}

\vskip -2.4in {\hfill \sl Submitted to The Astrophysical Journal Letters
1999 May 7}
\vskip 2.3in

\begin{abstract}

The nuclear rotation curve of M31, as observed by the {\it Hubble Space
Telescope\/} Faint Object Camera Spectrograph, shows a significant
disturbance coinciding with the off-center brightness peak, P1. This
$\pm 60 \kms$ feature is distinguished by
a local velocity maximum centered on P1 and a local minimum $\sim 0\farcs08$
closer to P2. If the M31 double nucleus is an eccentric disk with an
off-center density concentration, as suggested by Tremaine, then the
self-gravity of the disk can produce just such a disturbance. The expected
kinematic signature is calculated approximately by examining sequences of
closed periodic orbits in a Kepler potential perturbed by a model disk
potential that precesses at constant frequency. The perturbation 
forces a steep negative eccentricity gradient in the sequence of closed
orbits through the densest part of the disk, which reverses the arrangement of
periapsis and apoapsis with respect to the central mass.
Stars making up the inner part of the density concentration are at
apoapsis, while stars making up the outer part are at periapsis,
producing a steep local velocity gradient. This
result is independent of the details of the mass distribution. The
projected rotation curve of the model is shown to closely resemble that of M31,
giving strong support to the eccentric disk picture.

\end{abstract}


\section{Introduction\label{s.introduction}}

Currently viable explanations for the double-peaked structure of
the nucleus of M31 revealed by the {\it Hubble Space Telescope\/} (\hst;
Lauer et al. \markcite{Lau93}1993, 1998) center around two
basic scenarios: first, that the off-center brightness peak, P1,
represents a transient structure, possibly a star cluster on the verge of
tidal disruption (Emsellem \& Combes \markcite{EmC97}1997); second, that
P1 is an equilibrium configuration, resulting from a statistical accumulation
of stars near the apoapsides of their orbits in an eccentric ring or disk
(Tremaine \markcite{Tre95}1995, hereafter T95). Groundbased spectroscopy at
arcsecond (Bacon et al. \markcite{Bac94}1994) and better (Kormendy \& Bender
\markcite{KB99}1999, hereafter KB99) resolution shows asymmetries in the
rotation and dispersion profiles across the two brightness peaks. These
asymmetries are in the sense expected in the eccentric disk picture,
prompting KB99 to argue in strong support of the T95 model.
However, the sinking-cluster scenario has enough adjustable parameters
that, with sufficient perseverance, an adequate fit to the data
could probably be found.

The highest resolution kinematic data to date
come from the f/48 long-slit spectrograph of the \hst\ Faint
Object Camera (FOC; Statler et al. \markcite{Sta99}1999, hereafter SKCJ).
The FOC rotation curve is completely consistent with the groundbased data,
when the former is convolved to the resolution of the latter.
In addition, the FOC data show kinematic features at smaller scales. The
most significant is a disturbance to the rotation curve
in P1, superficially resembling a barely resolved local rotation in the
same sense as the overall rotation of the nucleus (fig.\ \ref{f.rotcurve}c
below). The FOC data are
limited by low signal-to-noise ($S/N$) ratio, and confirmation by
upcoming Space Telescope Imaging Spectrograph (STIS) observations is
certainly desirable. However, the ``P1 wiggle'' is found in the region
of highest $S/N$ in the FOC data, and the peak-to-peak amplitude
of $\sim 120 \kms$ is robust and insensitive to details of the
reduction process.

It is difficult to argue that the P1 wiggle could be a natural consequence of
the sinking cluster scenario. If P1 is a bound object, its luminosity
($\sim 10^6 \lsun$) and characteristic radius ($\sim 1 \pc$) would
suggest a severely tidally truncated, collapsed-core globular cluster.
But a rotation velocity $\sim 60\kms$ would correspond to
$V/\sigma > 1$, at least two times higher than observed in Galactic
globulars (e.g., Gebhardt et al.\ \markcite{Geb97}1997) or inferred
from the their flattenings (Davoust \& Prugniel \markcite{Dav90}1990).
Spin-up by an earlier tidal interaction with the central black hole
(BH) is conceivable, but it is hard to see how such an interaction could have
avoided disrupting the cluster.

On the other hand, a local distortion to the rotation curve of this magnitude
may be a
natural consequence of the eccentric disk picture. The shape of the distortion
suggests that the self-gravity of the disk may be at work. It is easy to
see how such a feature could arise in the simple case of a massive {\it
circular\/} ring centered on the BH. The ring pulls outward on objects
in its interior; thus the ring potential would lower the circular velocity
for orbits just inside the ring, and raise it for orbits just outside,
distorting the rotation curve in the sense observed. Of course, this distortion
would be seen symmetrically on both sides of the center.
In this {\it Letter\/}, I show that the self-gravity of an eccentric
disk will naturally produce the same kind of
feature in the rotation curve---though for a different reason---only
on one side, as is seen in the FOC data.

The plan of this {\it Letter\/} is as follows: In \S\ \ref{s.disks} I
construct a simple model for a non-self-gravitating eccentric disk, whose
parameters are consistent with the T95 model for M31. The purpose
of the initial model is only to provide a plausible density distribution,
from which I calculate a plausible perturbation to the otherwise Keplerian
potential. In \S\ \ref{s.orbits}, I examine the closed periodic
orbits in the perturbed potential, in a frame rotating at an assumed
precession speed $\Omega_p$; these orbits will be the parents of more general
quasi-periodic orbits that will be populated in the self-gravitating disk.
The character of the perturbation forces a steep negative eccentricity
gradient in the sequence of orbits moving outward through the
densest part of the disk. This gradient reverses the arrangement of periapsis
and apoapsis, such that stars making up the inner part of the density
concentration are at apoapsis, while stars making up the outer part are
at periapsis. Elementary considerations of celestial mechanics
show why this must be the case, independent of the details
of the mass distribution. While self-consistent models are outside the
scope of this {\it Letter\/}, I estimate in \S\ \ref{s.rotation} the effect
on the rotation curve by integrating the closed-orbit velocity field over an
aperture approximating the FOC slit and show that the expected signature
is qualitatively close to the observed rotation curve. Finally, I discuss
in \S\ \ref{s.discussion} the implications for more realistic models and the
prospects for using details of the rotation curve to constrain the masses
of the disk and BH.

\section{Mass Distribution in an Eccentric Disk\label{s.disks}}

To calculate a plausible density distribution, I assume a cold,
infinitesimally thin disk of stars on aligned elliptical Kepler
orbits. I place the BH of mass $M$ at the origin and the
common line of apsides along the $x$ axis in a cartesian system, with
periapsides at positive values of $x$. The eccentricity $e$ as a function of
semimajor axis $a$ in the initial model is specified by a fixed function
$e(a)$, which I assume changes sufficiently slowly that the orbits are not
mutually intersecting. The gravity of the bulge is ignored.

For a mass $dM$ distributed in a phase-independent way around a single
orbit, the mass per unit length $\ell$ would be given by $dM/vP$, where $v$ is
the instantaneous speed and $P$ the period. For a continuum of orbits
labeled by their semimajor axes $a$ and populated so that the mass per
unit interval of $a$ is $\mu(a)$, the mass in the area $da\,d\ell$
at $(a,\ell)$ is $dM = \mu(a) da\,d\ell / vP$. Replacing $a$ by length
$s$ measured perpendicular to the orbit introduces a factor $da/ds =
|\nabla a|$. Writing this factor explicitly in terms of the eccentricity law
$e(a)$, and using the standard formulae for Kepler orbits, I obtain for the
surface density of the disk, after some algebra,
\beq\label{e.surfacedensity}
\Sigma(a,x) = {\mu(a) \over 2 \pi a} {(1-e^2)^{1/2} \over 1 - e^2
	- (2ae + x)e^\prime}.
\eeq
In equation (\ref{e.surfacedensity}), $x$ is the cartesian coordinate and
$e^\prime \equiv de/da$. Note that having a density maximum
at apoapsis ($x = -a[1+e]$) requires that the eccentricity decrease outwards.

A particularly simple choice for $e(a)$ that
is qualitatively consistent with the behavior in the T95 model is
\beq\label{e.confocal}
e(a) = {\Delta \over a}.
\eeq
Equation (\ref{e.confocal}) implies that the orbits are all confocal,
sharing the focus at the origin occupied by the BH and the empty
focus at $x=-2\Delta$.
The surface density of a confocal disk takes on a simple form
in the elliptic coordinate system $(u,w)$ defined by
\beq\label{e.coordinates}
x = \Delta (\cosh u \sin w -1), \qquad y = \Delta \sinh u \cos w.
\eeq
In these coordinates each orbit follows an ellipse $u={\rm constant}$,
and the surface density is given by
\beq\label{e.density}
\Sigma(u,w) = {\mu (\Delta \cosh u) \over 2 \pi \Delta \cosh u}
	{\tanh u \over \cosh u + \sin w}.
\eeq
The potential can be computed by standard methods, most
efficiently by an expansion in suitable basis functions. For this work I
simply evaluate the integral
\beq\label{e.potential}
\Phi = -{G \over \Delta} \int_0^{2\pi} dw^\prime \int_0^\infty du^\prime
        {\Sigma(u^\prime,w^\prime)
        \left(\sinh^2 u^\prime + \cos^2 w^\prime\right) \over
        \left\{\left[\cosh(u-u^\prime) - \cos(w-w^\prime)\right]
        \left[\cosh(u+u^\prime) + \cos(w+w^\prime)\right]\right\}^{1/2}}.
\eeq
numerically and tabulate it on a grid in $(u,w)$.

\begin{figure}[t]{\epsfxsize=6.in\hfill\epsfbox{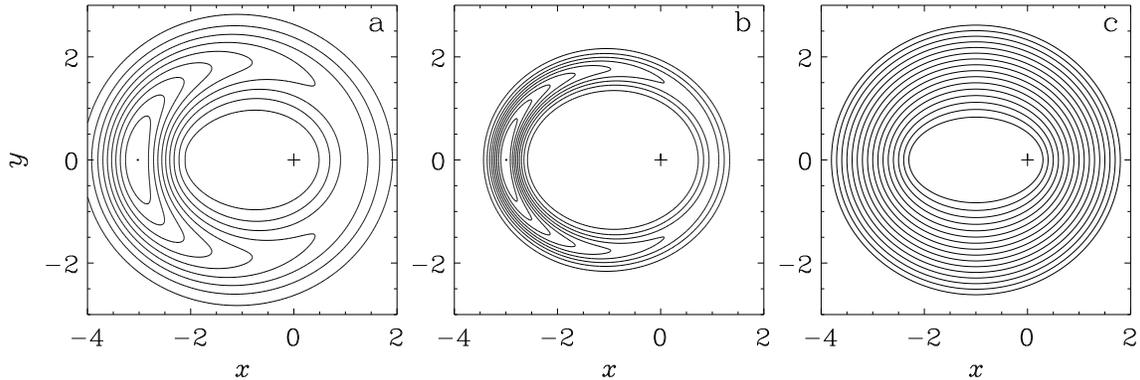}\hfill}
\caption{\footnotesize ($a$) Density contours for a ``wide'' eccentric disk
built from
confocal, aligned Kepler orbits, with the mass distribution as given in eq.
(\protect\ref{e.mu}) and width $\sigma=0.5$. Contours are at
$0.1,0.2,\ldots,1.0$ of the maximum density. Cross marks the central point
mass.  ($b$) Same as ($a$), but for the ``narrow''
disk, with $\sigma=0.2$. ($c$) Representative orbits in the disks,
following the eccentricity law in eq. (\protect\ref{e.confocal}).}
\label{f.density}
\end{figure}

A choice for $\mu(a)$ that produces a disk with a central density
minimum and a manageable outer cutoff is
\beq\label{e.mu}
\mu(a) = \mu_0(a-\Delta)\exp \left[ - {(a-a_0)^2 \over 2 \sigma^2} \right],
\eeq
where the leading factor prevents density singularities at the foci.
The numerical results below are computed in units where $G=M=\Delta=1$.
Taking $a_0=2$ places the peak density on orbits
with eccentricities close to $e=0.5$; for comparison, the innermost ringlet
in the T95 model, which contributes most of the density in P1, has
$e=0.44$. I consider two models with $\mu(a)$ as given
in equation (\ref{e.mu}) and $a_0=2$: a ``wide'' disk with $\sigma=0.5$,
and a ``narrow'' disk with $\sigma=0.2$. The center of mass is at
$x=-1.5$ for both disks. The surface densities
are shown in figure \ref{f.density}, along with some representative
orbits. I emphasize again that the purpose of these models is only to
provide a plausible perturbing potential.

\section{Periodic Orbits\label{s.orbits}}

Both the disk models discussed above and the T95 model are built entirely
from aligned, periodic orbits. A real self-gravitating disk with finite
velocity dispersion will be made predominantly from quasiperiodic orbits
whose parents are nearly elliptical, periodic orbits elongated in the
same sense as the disk. As long as the dispersion is not too large, the
kinematics of the quasiperiodic orbits will approximately follow that of
their periodic parents. Therefore a first approximation to the disk
kinematics can be found by examining the sequence of periodic orbits elongated
in the $x$ direction in the perturbed potential.
Consider the effect of the perturbation on purely Keplerian orbits. If
it were fixed in an inertial frame, the perturbing potential [equation
(\ref{e.potential})] would initially drive a precession of each
Kepler orbit at a frequency $\Omega(a,e)$, depending on semimajor axis and
eccentricity. Alternatively, if it were fixed in a frame rotating
at frequency $\Omega(a,e)$, then the perturbed orbit with that $a$ and 
$e$ would be closed in the rotating frame. I assume that the disk is stationary
in a frame rotating at a fixed precession speed $\Omega_p$; then the set of
orbits satisfying $\Omega(a,e)=\Omega_p$ are the closed orbits
in the disk.

I calculate sequences of periodic orbits for various assumed values of
$\Omega_p$ by direct integration of the equations of motion in a
frame rotating about the system barycenter. Orbits are launched
perpendicularly from the $x$ axis and the
initial velocities adjusted iteratively so that the next $x$ axis
crossing occurs with $v_x=0$. This procedure is not intended
to find all of the periodic orbits, only the nearly elliptical ones
elongated in the $x$ direction.

\begin{figure}[t]{\epsfxsize=3.in\hfill\epsfbox{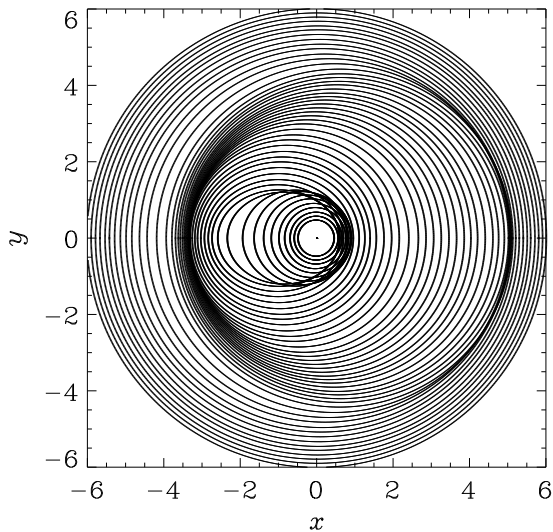}\hfill}
\caption{\footnotesize Closed periodic orbits in the wide disk, including
self-gravity and
uniform precession, for a disk mass $m=0.025$ and precession speed
$\Omega=0.006$. The radial variation of eccentricity is required for
uniform precession, and forces the crowding of orbits near the region of peak
density.}
\label{f.orbits}
\end{figure}

A typical sequence of orbits in the wide ($\sigma=0.5$) disk is
shown in Figure \ref{f.orbits}. This sequence is computed for a disk mass
$m=0.025$ and precession speed $\Omega_p=0.006$. All of the numerical
results below are computed using the same disk mass; to leading
order the results depend only on the ratio $m/\Omega_p$ and thus
can be scaled to other masses. Notice that: (1) the innermost and outermost
orbits are nearly circular; (2) a maximum eccentricity is reached by
orbits whose apoapsides are slightly inside the peak density in the disk; and,
most important,
(3) many orbits of rapidly declining eccentricity pass through nearly the same
apoapsis, in this case just outside the density peak.

\begin{figure}[t]{\epsfxsize=5in\hfill\epsfbox{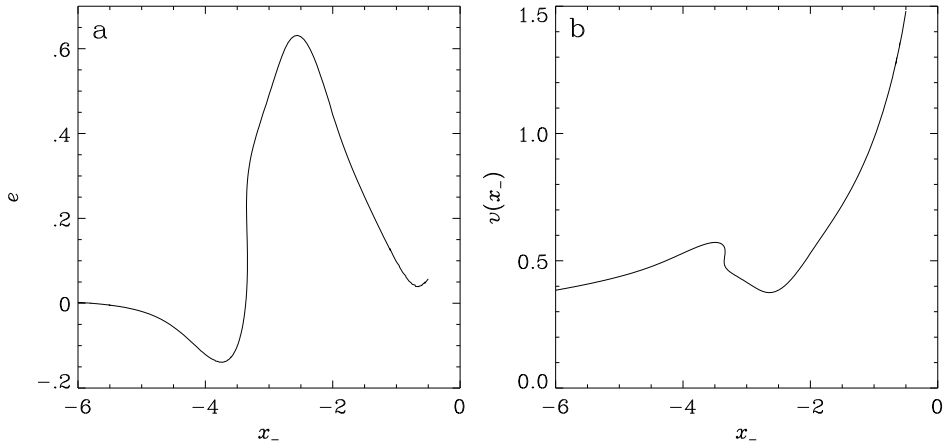}\hfill}
\caption{\footnotesize ($a$) Eccentricity of the orbits in
Fig.\ \protect\ref{f.orbits}
plotted against crossing point on the negative $x$ axis. The eccentricity
changes sign through the region of peak density, indicating that
periapsis and apoapsis change sides with respect to the center.
($b$) Local orbital speed crossing the negative $x$ axis plotted against
$x$ axis position. Velocities increase sharply outward just
outside the density peak (at $x=-3.1$); the plotted function is actually
double-valued owing to orbit crossing.}
\label{f.profiles}
\end{figure}

Since the perturbed orbits are not precisely elliptical, eccentricity is
defined here in terms of the points $x_+$ and $x_-$ where the orbit
intersects the positive and negative $x$ axis, respectively. I let
$e \equiv (x_- + x_+)/(x_- - x_+)$, so that $e>0$ ($e<0$) indicates periapsis
at positive (negative) $x$.
The run of $e$ against $x_-$ is shown in Figure \ref{f.profiles}a.
Notice that $e$ goes negative for orbits outside the density peak.
It is easy to see why $e$ has to change sign. Consider
an orbit just inside the density peak. The perturbation
can be approximated by an outward impulse applied to the orbit at apoapsis,
which induces a precession in the prograde direction (Brouwer \&
Clemence \markcite{Bro61}1961). Conversely, an
orbit just outside the density peak receives an inward impulse, which must be
applied at periapsis to cause precession in the same direction. The required
gradient, $e^\prime < 0$, is just what equation (\ref{e.surfacedensity})
says is needed to produce a mass concentration along the negative $x$ axis.
Note that equation (\ref{e.surfacedensity}) applies for $e$ of either
sign; thus in the region where $e<0$, the outward increase in $|e|$ still
contributes to the mass concentration at negative $x$.
This argument also shows that the disk precession must be prograde,
since retrograde precession would require $e^\prime > 0$, putting the mass
concentration on the wrong side to produce the needed
impulse.\footnote{Goldreich \& Tremaine \markcite{Gol79}(1979) apply a
similar argument the $\epsilon$ ring of Uranus. In that situation,
however, the ring self-gravity acts to {\it lower\/} the precession
rate driven by the Uranian quadrupole; hence the eccentricity increases
outwards and the densest part of the ring is at periapsis.}
At smaller and larger $a$, $|e|$ returns to near zero.
For these orbits, the perturbation approximates a constant force
in the negative $x$ and $r$ directions, respectively. In either case,
the precession rate contains a leading factor $(1-e^2)^{1/2}/e$
(Brouwer \& Clemence \markcite{Bro61}1961). Orbits more distant from
the density peak must therefore be less eccentric to maintain the same
precession rate.

Relatively simple celestial mechanics thus shows that $e$ must change sign (in
the convention used here, from positive to negative) in any
near-Keplerian disk with an eccentric density peak, in order for the disk
to precess
uniformly. The change in the sign of $e$ means that stars contributing to the
inner part of the density concentration are lingering near apoapsis, having
risen from smaller radii, while stars making up the outer part are swinging
through periapsis, having fallen from larger radii. This is the key to
the observable kinematic signature.

\section{Effect on the Rotation Curve\label{s.rotation}}

\begin{figure}[t]{\epsfxsize=3.1in\hfill\epsfbox{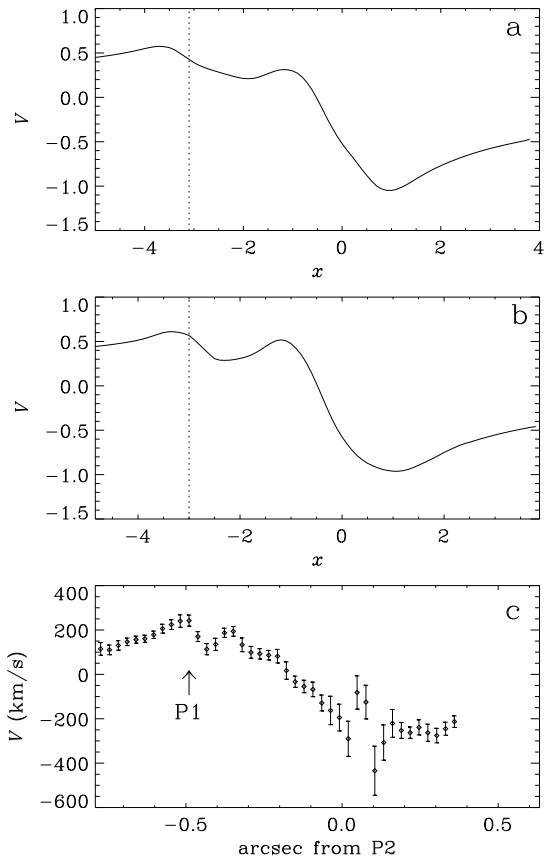}\hfill}
\caption{\footnotesize ($a$) Projected rotation curve for the wide disk,
with mass $m=0.025$
and precession speed $\Omega=0.004$. Contributions from the bulge are
ignored, and the velocity field is averaged over an aperture
approximating the FOC slit (see \S\ 4) laid along the disk major axis.
Vertical dotted line marks the location of the peak density. ($b$) As in
($a$), but for the narrow disk with $m=0.025$ and $\Omega=0.006$.
($c$) Central region of the FOC rotation curve of M31, from Statler et
al. (1999). Arrow marks the center of P1 and the location of the ``P1
wiggle,'' the form of which is closely reproduced by the models.}
\label{f.rotcurve}
\end{figure}

Figure \ref{f.profiles}b shows the velocities of orbits crossing the negative
$x$ axis as a function of the crossing point $x_-$. The velocity falls below
the local circular speed for $x_- > -3.4$ because of the positive
eccentricity, then abruptly rises as $e$ drops through zero.
The function $v(x_-)$ is actually double-valued because the orbits are
mutually crossing near apoapsis in the region where $e$ is changing rapidly
(Fig.\ \ref{f.orbits}).
Orbit crossing occurs for both of the disk density models considered here,
though it is reduced somewhat by higher precession speeds. The rather rapidly
precessing case shown in Figures \ref{f.orbits} and \ref{f.profiles} was
chosen more for clarity than for realism. In reality, the precession speed
will be set by self-consistency. A reasonable estimate for
$\Omega_p$ can be obtained by requiring that the orbit through the
density maximum of the original confocal model precess with the
disk, retaining its original eccentricity. This gives slightly slower
speeds of $\Omega_p=0.004$
for the wide disk and $\Omega_p=0.006$ for the narrow disk, for a
mass of $0.025$.

The double-valued nature of the rotation curve in Figure \ref{f.profiles}b
will, of course, be washed
out by projection effects. To approximate the observable
signature, I project the disk velocity fields for a line of sight
parallel to the $y$ axis, adopting for the disk density that of the
original confocal model. To compare with the FOC rotation curve, I let
the distance from the origin to the density maximum in the model
correspond to the P1-P2 separation of $0\farcs 49$. In the T95 model,
the disk has an inclination of $77\arcdeg$, at which the FOC slit width
of $0\farcs 063$ projects to a width of $0\farcs 28$ in the disk plane.
This corresponds to a band of width $\Delta y=1.8$ about the $x$
axis, over which I integrate the line of sight velocity. Since I
completely ignore any contribution from the bulge, either to the light
or to the rotation, the results should not be taken too literally,
especially near the center.

Figure \ref{f.rotcurve}a shows the projected rotation curve for the
wide disk with $m=0.025$ and $\Omega_p=0.004$. The dip and outer bump
produced by the steep part of the ellipticity profile are preserved in
projection. The near-Keplerian profile at small $x_-$ is smoothed into
an inner bump by the central hole in the density distribution and by the
finite slit width. The combination of these effects produces a wiggle
similar to that in the M31 rotation curve (fig.\ \ref{f.rotcurve}c);
note that there is no comparable feature on the anti-P1 side of the
center in either model or data. For the wide disk,
the model wiggle is quite a bit too
broad, a discrepancy remedied somewhat by narrowing the mass
distribution. Figure \ref{f.rotcurve}b shows the rotation curve for the narrow
disk with the same mass and $\Omega_p=0.006$. Though the fit is
still not exact, the wiggle is noticeably narrowed; in particular, the
outer bump now closely coincides with the density peak, in agreement
with the FOC data. The results in Figure \ref{f.rotcurve} are not
qualitatively changed by $\pm 50\%$ variations in the assumed value of
$\Omega_p$.

\section{Discussion\label{s.discussion}}

I have shown that in an eccentric near-Keplerian disk with an off-center
density concentration, the disk's self-gravity will affect
the periodic orbits in such a way as to produce an observable signature in
the projected rotation curve. This signature is a ``wiggle'' extending
through the region of peak density, with a local velocity maximum at or
just outside the peak, and a dip in the rotation curve just inside.
Details interior to this region depend on the
density structure, viewing geometry, and resolution. For a mass distribution
resembling that of the T95 model for the M31 double nucleus, the computed
wiggle closely resembles the observed FOC rotation
curve through P1. This agreement gives strong support to the basic
correctness of the eccentric disk picture.

A better understanding of the dynamics of M31's nuclear region should
lead to more accurate mass determinations for the central BH.
An enticing notion is that the P1 wiggle could be used to
constrain the disk-to-BH mass ratio, and thereby the mass of the BH by
way of an assumed mass-to-light ratio for the disk. Unfortunately, 
this is easier said than done.
Disk models with the same value of $m/\Omega_p$ produce the same
eccentricity law and rotation curve, to leading order.
Constraining the mass would thus require an independent estimate of the
precession speed. The T95 model has a disk mass of $0.16$, which would
imply $\Omega_p \sim 0.03$, by the arguments of \S\ 4. Approximating
the potential as Keplerian, this would put an inner Lindblad resonance
(ILR) outside the dense part of the disk, in the vicinity of
$r \sim 1\arcsec$. But whether one should expect to detect an ILR in a
predominantly hot stellar system is, to say the least, problematical.

More realistic disk models will be needed both to obtain an
accurate mass for the central BH and to understand the stellar dynamics
of the nucleus. The simple models presented here neglect some
important complications. The $e(a)$ profile in Figure \ref{f.profiles}a
for the periodic orbits
would imply a much more sharply peaked density structure than assumed in
the confocal models in \S\ 2. But equation (\ref{e.surfacedensity}) for
the surface density is no longer valid when the orbits intersect,
which would appear to be a necessary consequence of self-gravity.
In reality, the radial width of the disk will be determined by
the velocity dispersion. A self-consistent disk will
mainly be composed of quasi-periodic orbits librating about the
closed orbits considered here.\footnote{Examples of these loop-like orbits
and their possible role in nuclear disks have recently been discussed
by Sridhar \& Touma \markcite{Sri99}(1999).} Sufficiently high dispersion
could wash out the kinematic signature of the periodic orbits;
self-consistent models will be required to determine at what dispersion
this occurs. If the P1 wiggle is a sign of disk self-gravity, the challenge
may really be to explain why M31's disk is dynamically
cold enough for the effect to be visible.

\acknowledgments

This work was supported by NSF CAREER grant AST-9703036. The author is
grateful to Ivan King, Joe Shields, Scott Tremaine, and Steve Vine for
helpful comments.

\end{document}